# ESTIMULACIÓN ELÉCTRICA TRANSCRANEAL: UNA BREVE INTRODUCCIÓN

M. Sc. Carlos G. Tarazona*, Md. Carlos E. Guerra**, Laura Chávez*** y Sebastián Andrade***

*Departamento de ciencias Básicas, Docente /UMB, Bogotá, Colombia
**Médico Cirujano / Universidad Nacional de Colombia, Bogotá, Colombia
*** Facultad de Ing. Biomédica, Estudiante /UMB, Bogotá, Colombia

e-mail: caragomezt@unal.edu.co

**RESUMEN:** *El objetivo principal de la estimulación eléctrica del cerebro es generar potenciales de acción a partir de la aplicación de campos electromagnéticos. Dentro de las técnicas disponibles, la estimulación eléctrica transcraneal (TES) representa un método popular de administración, el cual tiene la ventaja de no ser invasivo y económicamente más asequible. Este artículo busca ser una breve introducción para comprender en que consiste la TES desde el punto de vista de la física involucrada en ella y de algunos resultados relevantes al realizar la implementación de dicha técnica.*

**PALABRAS CLAVES:** *Cerebro, Conductividad Eléctrica, Estimulación Eléctrica, Estimulación Eléctrica transcraneal.*

**ABSTRACT:** **The main objective of the electrical stimulation of the brain is to generate action potentials from the application of electromagnetic fields. Among the available techniques, transcranial electrical stimulation (TES) represents a popular method of administration that has the advantage of being non-invasive and economically more affordable. This article aims to briefly introduce the reader into the understanding of TES in terms of the physics involved as well as for some of the relevant results of studies applying this technique.**

## 1. INTRODUCCIÓN

El cerebro adulto es un órgano con una masa aproximada de 1.4kg para una persona de una masa de 70kg, lo cual representaría solamente un 1.7% de la masa total.[1]

$$M_{cerebro} = (0.08 - 0.09)M^{0.66} \qquad (1)$$

Sin embargo, es impactante que esta masa relativamente pequeña sea capaz de controlar todas las funciones vitales y al mismo tiempo entender y responder al contenido de este texto. Así entonces, el estudio del cerebro desde el punto de vista fisiológico y físico es uno es uno de los aspectos más interesantes, no sólo debido a la increíble complejidad de su función y a las implicaciones de las terapias presentes y futuras, sino también a la física involucrada en éste.

El presente artículo pretende mostrar de una forma sencilla y clara algunos aspectos básicos de la física del cerebro, así como una introducción a la estimulación cerebral, en especial la estimulación eléctrica transcraneal TES (transcranial electrical stimulation). La TES es una técnica de particular interés en aplicaciones en el ámbito médico, tanto por su potencial asequibilidad como por no requerir métodos de administración invasivos.

## 2. ESTIMULACIÓN ELÉCTRICA DEL CEREBRO

En términos generales, la estimulación eléctrica del cerebro se refiere a la generación de potenciales de acción neuronales a partir de la aplicación de una corriente eléctrica (o carga eléctrica) en el tejido cerebral.[2] Se sabe que la estimulación eléctrica tiene un efecto principal sobre los axones o prolongaciones nerviosas, más que en el cuerpo de la neurona en sí mismo.[2] Los potenciales resultantes son experimentalmente medibles, y consisten tanto en la estimulación directa producida por el electrodo que entrega la corriente eléctrica, como en el reclutamiento de descargas que facilitan transmisiones neuronales posteriores (i.e. ondas D y ondas I). Así, las respuestas motoras generadas por impulsos de corta duración y alta intensidad tienen umbrales menores para músculos en contracción activa que en aquellos en reposo.[2]

Desde el punto de vista histológico y celular, la aplicación de corriente directa sobre el tejido neuronal ha mostrado generar cambios tanto en la forma de los cuerpos celulares como en la orientación y crecimiento de las prolongaciones nerviosas de neuronas, microglia y astrocitos.[3] Esta evidencia sugiere que la estimulación eléctrica no sólo genera impulsos eléctricos, sino que puede modificar la función del tejido estimulado.

También se ha observado que la presencia de un campo eléctrico externo es capaz de aumentar la permeabilidad de la membrana celular. Este fenómeno se ha denominado electroporación o electropermeabilización .[4] Sugar y Neumann han explicado previamente que estos estímulos eléctricos originan cambios en la distribución y orientación de los componentes de la bicapa lipídica de la membrana, formando un "poro" transitorio que puede variar en su tamaño y duración, pero que en esencia es reversible.[5]

Existen varios métodos de estimulación eléctrica cerebral, que según su forma de administración se clasifican en invasivos y no invasivos. Dentro de los primeros, la estimulación cerebral profunda (ECP, o DBS por sus siglas en inglés)[6] ha sido uno de los métodos más estudiados. La ECP funciona a través de uno o más electrodos ubicados en regiones específicas del cerebro, en contacto directo con éste mismo, los cuales reciben la energía aplicada de una batería externa (usualmente ubicada en el pecho), formando un circuito cerrado.[7] Esto implica realizar por lo menos dos incisiones y por ende es completamente invasiva y de mayor riesgo, además que plantear una redistribución de los electrodos para estimular una nueva área, planea una nueva intervención. Sin embargo, cuenta a su favor con el hecho de que la estimulación eléctrica puede ser más predecible, modelable y eficiente.[7]

Por otro lado, los métodos no invasivos utilizan electrodos fuera de la cavidad craneal, más exactamente sobre el cuero cabelludo, lo cual permite un ajuste de su ubicación de estos sin ningún inconveniente. Los electrodos son colocados con un gel que mejora la conductividad en su camino hacia el cerebro. Así, la estimulación eléctrica no invasiva del tejido cerebral provoca una redistribución de cargas y dipolos en un medio de alguna manera acuoso y heterogéneo[8]. Dado que el cerebro se puede considerar como un medio dieléctrico (el tejido cerebral en su mayoría está compuesto por agua, la cual es una molécula polar), un campo eléctrico externo aplicado a éste producirá un momento dipolar en el tejido debido al reacomodamiento de las moléculas, para oponerse a la dirección del campo externo.

Como resultado del desplazamiento de cargas y de reorientación de los momentos dipolares, la estimulación transcraneal permite la creación de campos eléctricos que dependen de la distribución de los electrodos sobre el cráneo, la cantidad de carga de cada uno de ellos y el área de interés dentro del cerebro.[9]

A diferencia de la ECP, una ventaja clara de la estimulación transcraneal resulta del hecho de ser un procedimiento no invasivo y que no requiere de ningún entrenamiento quirúrgico para su aplicación.[10] No obstante, esto sucede al costo de una mayor resistencia eléctrica de los tejidos circundantes. Por sencillez podemos considerar un modelo físico con tres capas principales: el cuero cabelludo, el cráneo y la sustancia gris, las cuales tienen unas resistencias respectivas de $2.22\Omega$, $177\ \Omega$ y $2.22\Omega$, y unas conductividades respectivas de $0.45045\ \Omega^{-1}$, $5.6{\times}10^{-3}\ \Omega^{-1}$ y $0.45045\ \Omega^{-1}$ [11]. De esta manera, al colocar un campo eléctrico externo, éste se encontrara "apantallado", debido a la alta resistencia del cráneo con respecto a las capas que lo rodean, lo cual implica la necesidad de utilizar campos eléctricos de más elevados, produciendo en consecuencia posibles molestias sobre el cuero cabelludo. Esta limitación ha estimulado el uso de otro tipo de técnicas, que al aplicar otro tipo de campos (i.e. magnéticos) se pueden evitar estas molestias potencialmente desagradables.[2]

Otros trabajos han reportado una amplia variabilidad en las propiedades resistivas del hueso (dependiendo del método usado y la composición del tejido), con valores de hasta $10,000\ \Omega$ cm para volúmenes óseos en regiones de alta resistividad como la cóclea humana.[12] Para el cráneo humano, los valores de conductividad también se han reportado en el rango de $32\ mS/m$ a $80\ mS/m$.[13] Muchos de los estudios de permitividad y conductividad en el cerebro han basado sus conclusiones en experimentos animales o en tejidos humanos con más de 24 horas postmortem, lo cual limita la extrapolación de los datos encontrados. Para este fin, Schmid et al. realizaron la medición de la permitividad y la conductividad en 20 cerebros humanos durante las primeras 10 horas postmórtem. Se encontraron valores de conductividad entre 1.13 y 2.09 S/m y valores de permitividad relativa aproximadamente entre 58 Y 54 para distintas frecuencias de estimulación de este órgano.[14] Los cambios de temperatura corporal debido al estado postmórtem, tienen un efecto directo en los valores de resistividad. Algunos investigadores sugieren también que la disminución de las propiedades dieléctricas de los tejidos podría obedecer a cambios en la conductividad iónica del tejido, así como a las consecuencias estructurales y metabólicas de la muerte del tejido. [14]

La tabla 1 nos permite hacernos una idea de qué tan variable es la resistividad dependiendo del material y del tipo específico del material:

**Tabla 1: Resistividades de diferentes materiales conductores (CI), semiconductores (SC) y conductores iónicos (CI) [1]**

| Tipo | Material | Resistividad (Ωm) |
|------|----------|-------------------|
| C | Plata | $1.5{\times}10^{-8}$ |
| SC | Carbono(grafito) | $3.5{\times}10^{-5}$ |
| A/CI | Membrana celular | $10^{6}$-$10^{9}$ |
| CI | Axoplasma | 2 |
| CI | Agua ultra pura | $10^{5}$ |
| CI | Agua uso domestico | 10-$10^{2}$ |

Otro aspecto importante a considerar es la anisotropía, que consiste en la variación de las propiedades físicas descritas según la dirección en la que son examinadas. En el cerebro, la sustancia gris es la principal responsable. La anisotropía cerebral puede explicarse por las diferencias en la morfología neuronal y la configuración de las redes nerviosas en cada región anatómica.[15] Esto hace que dependiendo de la forma como sea aplicado el campo eléctrico, podemos observar distintos efectos, siendo éstos más o menos perceptibles desde el punto de vista físico.

Como podemos ver, el análisis de los campos y potenciales eléctricos dentro del cerebro, debe considerar la existencia de al menos tres capas con distintas propiedades resistivas y la anisotropía cerebral, este conjunto de factores representa un reto en el modelamiento físico para este tipo de terapias. No obstante, el modelar los diversos tipos de terapias teóricamente, puede proporcionar información significativa para el avance científico en este campo.

### 3. Estimulación Eléctrica Transcraneal (TES)

La estimulación eléctrica transcraneal (TES) ha recibido una atención creciente en los últimos años, siendo así una de las principales representantes de los métodos no invasivos de estimulación cerebral. Desde 1994, la Federación Internacional de Neurofisiología Clínica ha realizado revisiones del estado del arte y las bases para la investigación en la estimulación eléctrica y electromagnética cerebral no invasiva. Su actualización más reciente provee mecanismos importantes para estos tipos de estimulación no invasiva, pero se debe resaltar que el enfoque principal está en los métodos de estimulación magnética, dado que no generan sensaciones incómodas, a diferencia de estímulos eléctricos en el cuero cabelludo con técnicas de TES con voltajes relativamente mayores.[2]

A través de la electroporación, Corovic y colaboradores han estudiado los valores de campo eléctrico para los cuales la membrana puede mantener sus propiedades físicas y biológicas, mientras se logra la introducción de sustancias en la célula. Para valores superiores al del campo eléctrico crítico, la célula puede sufrir daños irreversibles, lo cual también puede representar una alternativa en el tratamiento oncológico[16]. Otros estudios han apoyado esta técnica para la introducción de segmentos específicos de ADN en las células tratadas.[17, 18] Asimismo, el modelamiento de los campos eléctricos puede proveer información importante acerca de los efectos tumorales de TES en la terapia contra el cáncer. Usando un modelo experimental de glioblastoma (el tipo de tumor maligno primario más frecuente del cerebro), Wenger et al. encontraron que al aplicar un campo externo a una frecuencia de 200KHz y una amplitud de (1-3) V·cm$^{-1}$, se inhibía la proliferación de la célula maligna. Ellos también notaron la existencia de una orientación privilegiada del campo, en la cual el campo eléctrico debe estar de forma paralela al eje de la división celular para lograr su efecto. Bajo estas condiciones, después de 24 horas de aplicar el campo eléctrico en roedores, se encontró que la mitosis de las células cancerígenas estaba completamente interrumpida. Dado que el eje no tiene una orientación única, al aplicar el campo eléctrico en dos direcciones, la mitosis se podría detener más rápidamente[19].

La capacidad de modificar los patrones de excitabilidad neuronal sugiere que la TES tiene un amplio rango de aplicaciones en la medicina. En un estudio aleatorizado, 9 individuos obesos fueron aleatoriamente para recibir estimulación cerebral transcraneal de corriente directa en un medio ambiente controlado y bajo un período de 9 días. En comparación con los individuos que recibían una terapia placebo, los individuos tratados con estimulación transcraneal mostraron un consumo significativamente menor de calorías. Así, la modificación de la actividad neuronal por medio de la estimulación transcraneal prefrontal podría ser útil para el tratamiento de enfermedades como la obesidad, de tanta prevalencia en el mundo moderno.[20]

Otro aspecto interesante de la utilización de TES tiene que ver con las opciones en el desarrollo del aprendizaje, como fue propuesto por Krause y Cohen Kadosh.[21] Mediante la utilización de un estimulador no invasivo, los autores proponen un método para suplir deficiencias de aprendizaje en poblaciones y en pacientes con trastorno de déficit de atención e hiperactividad.[21] Otros autores han notado mejorías significativas en el desempeño para tareas de memoria y lenguaje cuando son estimulados con TES de corriente directa anodal.[22]

### 4. DISCUSIÓN

Desde hace al menos cuatro décadas, el cerebro ha sido objeto de estudio constante en el campo de la estimulación eléctrica. Como resultado, las aplicaciones actuales de estas terapias son muy diversas, desde el estudio de los efectos en la corteza motora hasta el tratamiento de enfermedades neurológicas y el cáncer.[2, 7, 19]

La estimulación eléctrica cerebral logra sus efectos mediante la generación de potenciales de acción que se propagan a través de vías neuronales definidas. Esta estimulación tiene lugar en los axones y no en los cuerpos neuronales, principalmente debido a que el soma neuronal tiene umbrales de activación más elevados. [2] Adicionalmente, estos potenciales pueden variar de forma significativa dependiendo del estado de estimulación neuronal previo.[2] En otras palabras, resulta mucho más fácil estimular una región previamente estimulada que una región en reposo. Además, otros cambios histológicos significativos tienen lugar en el proceso descrito, e implican otras consecuencias biológicas en el tejido estimulado.[3]

---

1.    Las notas de pie de página deberán estar en la página donde se citan.  Letra Times New Roman de 8 puntos

La administración de corrientes electromagnéticas al cerebro puede ser invasiva o no invasiva. Hemos encontrado que los métodos invasivos pueden ser más predecibles y controlados, mientras que los métodos no invasivos pierden su especificidad en la necesidad de incrementar el estímulo eléctrico como resultado de los tejidos que debe atravesar antes de llegar al cerebro. No es una sorpresa entonces que los métodos de estimulación magnética hayan tenido una mayor aceptación para muchos investigadores. Sin embargo, la facilidad de desarrollar la estimulación eléctrica extracraneal ha hecho el método muy popular y de muy rápida difusión.

Por otro lado, la necesidad de realizar procedimientos quirúrgicos o invasivos para implantar neuroestimuladores representa un riesgo elevado para los pacientes que se benefician de estimulación cerebral profunda, en comparación con aquellas condiciones en las cuales el uso de electrodos externos resulta más económico y requiere de menor entrenamiento para el personal que potencialmente debería administrarla.

Finalmente, la distribución de los electrodos en la estimulación eléctrica no invasiva tiene la capacidad de generar potenciales eléctricos, los cuales a su vez afectan la distribución de cargas y pueden facilitar o impedir la despolarización neuronal en unas áreas o en otras. Resulta entonces de especial importancia entender más a fondo los resultados de dichas terapias desde el punto de vista de la física aplicada. Por ende, hay una necesidad imperante en conocer el modelamiento teórico de los eventos físicos que tienen lugar con la TES, y así poder explicar los resultados alentadores, en especial los relacionados con la neuromodulación.[22]

# 5. CONCLUSIONES

La producción de corrientes inducidas o campos eléctricos inducidos debido a la colocación de electrodos fuera de la cavidad craneana    tienen un efecto significativo y medible en las respuestas neuronales de la región estimulada. Dentro de las diversas modalidades de estimulación existentes, la estimulación eléctrica transcraneal representa una alternativa muy asequible y relativamente segura, si lo comparamos con los métodos de estimulación cerebral profunda y con los métodos de estimulación magnética que son significativamente más costosos. Como resultado de una revisión corta de la literatura, hemos entendido la necesidad que representa comprender los aspectos físicos de las interacciones de campos eléctricos externos en los potenciales neuronales, dado que éstos solo se han estudiado tangencialmente y en función de los resultados obtenidos de forma experimental, y no partiendo de un modelamiento físico de base.

# 6. BIBLIOGRAFÍA

[1]     F. Cussó, C. López, y R. V. Lázaro, *Física de los procesos biológicos*: Ariel, 2004.

[2]     P. M. Rossini, D. Burke, R. Chen, L. G. Cohen, Z. Daskalakis, R. Di Iorio, *et al.*, "Non-invasive electrical and magnetic stimulation of the brain, spinal cord, roots and peripheral nerves: Basic principles and procedures for routine clinical and research application. An updated report from an I.F.C.N. Committee," *Clin Neurophysiol*, vol. 126, pp. 1071-107, Jun 2015.

[3]     S. J. Pelletier, M. Lagace, I. St-Amour, D. Arsenault, G. Cisbani, A. Chabrat, *et al.*, "The morphological and molecular changes of brain cells exposed to direct current electric field stimulation," *Int J Neuropsychopharmacol*, vol. 18, Mar 2015.

[4]     E. Neumann, M. Schaefer-Ridder, Y. Wang, and P. H. Hofschneider, "Gene transfer into mouse lyoma cells by electroporation in high electric fields," *Embo j*, vol. 1, pp. 841-5, 1982.

[5]     I. P. Sugar and E. Neumann, "Stochastic model for electric field-induced membrane pores. Electroporation," *Biophys Chem*, vol. 19, pp. 211-25, May 1984.

[6]     U. Silverthorn Dee, "Fisiología humana, un enfoque integrado," *Médica Panamericana. 4ª edición. Bs As*, 2008.

[7]     D. T. Brocker and W. M. Grill, "Chapter 1 - Principles of electrical stimulation of neural tissue," in *Handbook of Clinical Neurology*. vol. Volume 116, M. L. Andres and H. Mark, Eds., ed: Elsevier, 2013, pp. 3-18.

[8]     K. R. Foster, J. L. Schepps, R. D. Stoy, and H. P. Schwan, "Dielectric properties of brain tissue between 0.01 and 10 GHz," *Phys Med Biol*, vol. 24, pp. 1177-87, Nov 1979.

[9]     A. Opitz, W. Paulus, S. Will, A. Antunes, and A. Thielscher, "Determinants of the electric field during transcranial direct current stimulation," *Neuroimage*, vol. 109, pp. 140-50, Apr 1 2015.

[10]    G. A. Castillo Castelblanco, E. García Cossio, and S. Castillo Peñuela, "Synchronization with transcranial brain stimulation," *Acta Neurológica Colombiana*, vol. 30, pp. 103-107, 2014.

[11]    J. I. Padilla Buriticá, "Localización de focos epilépticos mediante el análisis de registros EEG basada en modelos paramétricos y separación ciega de fuentes= EEG source localization based on parametric models and blind source separation," Universidad Nacional de Colombia-Sede Manizales, 2011.

[12]    T. K. Malherbe, T. Hanekom, and J. J. Hanekom, "The effect of the resistive properties of bone on neural excitation and electric fields in cochlear implant models," *Hear Res*, vol. 327, pp. 126-35, Sep 2015.

[13]    R. Hoekema, G. H. Wieneke, F. S. Leijten, C. W. van Veelen, P. C. van Rijen, G. J. Huiskamp, *et al.*, "Measurement of the conductivity of skull, temporarily removed during epilepsy surgery," *Brain Topogr*, vol. 16, pp. 29-38, Fall 2003.

[14]    G. Schmid, G. Neubauer, and P. R. Mazal, "Dielectric properties of human brain tissue measured less than 10 h postmortem at frequencies from 800 to 2450 MHz," *Bioelectromagnetics*, vol. 24, pp. 423-30, Sep 2003.

[15]    M. B. Hansen, S. N. Jespersen, L. A. Leigland, and C. D. Kroenke, "Using diffusion anisotropy to characterize neuronal morphology in gray matter: the orientation distribution of axons and dendrites in the NeuroMorpho.org database," *Front Neurosci*, vol. 7, p. 31, 2013.

[16]    S. Corovic, M. Pavlin, and D. Miklavcic, "Analytical and numerical quantification and comparison of the local electric field in the tissue for different electrode configurations," *Biomed Eng Online*, vol. 6, p. 37, 2007.

[17]    J. Szczurkowska, M. dal Maschio, A. W. Cwetsch, D. Ghezzi, G. Bony, A. Alabastri, *et al.*, "Increased

performance in genetic manipulation by modeling the dielectric properties of the rodent brain," *Conf Proc IEEE Eng Med Biol Soc,* vol. 2013, pp. 1615-8, 2013.

[18]     S. Corovic, I. Lackovic, P. Sustaric, T. Sustar, T. Rodic, and D. Miklavcic, "Modeling of electric field distribution in tissues during electroporation," *Biomed Eng Online,* vol. 12, p. 16, 2013.

[19]     C. Wenger, R. Salvador, P. J. Basser, and P. C. Miranda, "The electric field distribution in the brain during TTFields therapy and its dependence on tissue dielectric properties and anatomy: a computational study," *Phys Med Biol,* vol. 60, pp. 7339-57, Sep 21 2015.

[20]     M. E. Gluck, M. Alonso-Alonso, P. Piaggi, C. M. Weise, R. Jumpertz-von Schwartzenberg, M. Reinhardt, *et al.,* "Neuromodulation targeted to the prefrontal cortex induces changes in energy intake and weight loss in obesity," *Obesity (Silver Spring),* vol. 23, pp. 2149-56, Nov 2015.

[21]     B. Krause and R. Cohen Kadosh, "Can transcranial electrical stimulation improve learning difficulties in atypical brain development? A future possibility for cognitive training," *Dev Cogn Neurosci,* vol. 6, pp. 176-94, Oct 2013.

[22]     E. K. Hussey, N. Ward, K. Christianson, and A. F. Kramer, "Language and Memory Improvements following tDCS of Left Lateral Prefrontal Cortex," *PLoS One,* vol. 10, p. e0141417, 2015.

---

1.     Las notas de pie de página deberán estar en la página donde se citan.  Letra Times New Roman de 8 puntos